# Accuracy in Spreadsheet Modelling Systems

Thomas A. Grossman
OPMA, Haskayne School of Business, Calgary, Alberta, Canada T2N 1N4
Thomas.Grossman@Haskayne.UCalgary.Ca

**ABSTRACT**

Accuracy in spreadsheet modelling systems can be reduced due to difficulties with the inputs, the model itself, or the spreadsheet implementation of the model. When the "true" outputs from the system are unknowable, accuracy is evaluated subjectively. Less than perfect accuracy can be acceptable depending on the purpose of the model, problems with inputs, or resource constraints. Users build modelling systems iteratively, and choose to allocate limited resources to the inputs, the model, the spreadsheet implementation, and to employing the system for business analysis. When making these choices, users can suffer from expectation bias and diagnosis bias. Existing research results tend to focus on errors in the spreadsheet implementation. Because industry has tolerance for system inaccuracy, errors in spreadsheet implementations may not be a serious concern. Spreadsheet productivity may be of more interest.

## 1. INTRODUCTION

Research on spreadsheet accuracy tends to focus on errors in spreadsheet outputs, where errors are defined as deviations from known correct outputs. However, many spreadsheets implement models of d business situation. These models necessarily simplify complexity and selectively ignore certain aspects of the problem. Creating such models is a difficult art. Models are inherently subjective, and two people with equivalent domain knowledge can devise different models. Therefore, the "correct" outputs of a spreadsheet may not be known, and may not even be knowable. This complicates any discussion of spreadsheet errors.

Additional complications arise because the data that are input to a spreadsheet model can be less than ideal. Inputs can have errors, both known and unknown. Some data are inherently uncertain, and must be modelled as probability distributions or be treated as point estimates.

To understand how difficulties with models and data inputs interact with spreadsheet errors, I examine "spreadsheet modelling systems" where a *spreadsheet* implementing a *model* computes outputs based on inputs. Inaccuracies in system outputs can occur in the inputs, the model, and of course in the spreadsheet implementation. The user has limited control over these three sources of inaccuracy, and must divide his limited development resources among them and also to the task of using the system for analysis.

I start by formalizing what I mean by a "model" and a "spreadsheet modelling system". I then discuss accuracy, explain the difficulties in evaluating accuracy for spreadsheet modelling systems, and why "perfection" can be an inappropriate standard of accuracy. I discuss how each element of the system can be a source of inaccuracy. I argue that as the user struggles to balance the sources of inaccuracy, he may rationally choose to incur spreadsheet error. I discuss how a user's judgmental biases can affect the accuracy of the system. I conclude with observations regarding existing research, and industry's (lack of) reaction to that research.

## 2. SPREADSHEET MODELING SYSTEMS

Spreadsheets programs became popular because they provided managers a dramatically better way to create and compute business models then the previous alternative of mainframe software created by an independent information technology department. If (or perhaps when) someone invents a technology that is dramatically better than spreadsheets for creating and computing business models, spreadsheet usage will likely decline rapidly.

A **model** is a representation of a system that can be used to predict performance of some aspect of a business situation. As illustrated in Figure 1, a model converts inputs to outputs. A **business model** is set of specifications for using inputs to generate quantitative outputs for some business situation. I use "model" to refer only to business models (as opposed to physical models, social models, engineering models, etc.).

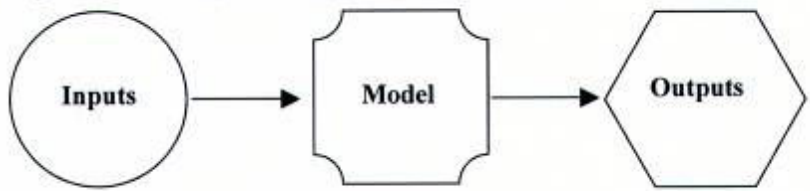

Figure 1,    Input-Output View of a Model

A model can be represented verbally (such as a software specification), in algebraic notation, in a spreadsheet, or as an algorithm (including a procedural computer program). As an example, consider a simple profit model, represented algebraically:

$$\text{Profit} = \text{Revenue} - \text{Cost}$$

This model is illustrated in Figure 2. If Revenue input is 1000 and the Cost input is 900, the Profit output is 1000 - 900 = 100. It is trivial to implement this model on a calculator, in a spreadsheet, or in a procedural programming language.

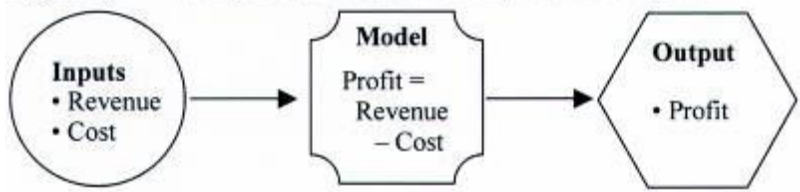

Figure 2,    Simple Profit Model with Inputs and Outputs

In general, the *model* is the core intellectual asset and software being is used to represent and compute a model. In business, the software of choice for models is almost always a spreadsheet. It is important to remember that a spreadsheet is a means to an end, not an end in itself.

The process of creating a model is called **modelling.** Modelling is the specification and use of a model. Modelling is inherently personal, creative, and subjective. Unfortunately, little is known about modelling. The scant existing research focuses on expert modellers in the field of operational research. The research show that modelling is an iterative process, and that modelling is viewed as an "art" that is difficult to study and teach [Powell 1995 1998, Willemain 1994a 1994b].

Spreadsheets are a uniquely powerful platform for modelling by ordinary business people. Spreadsheets have changed business modelling from an exclusive activity performed by elite operational researchers into a routine business activity. There is virtually no research on "end user modelling", which is modelling by ordinary businesspeople. However, spreadsheet users often do not distinguish between the act of modelling, and the act of programming a model in a spreadsheet [Grossman 2002].

Figure 3 extends Figure 1 to a *spreadsheet modelling system* that uses spreadsheet software to implement a model that convert inputs to outputs.

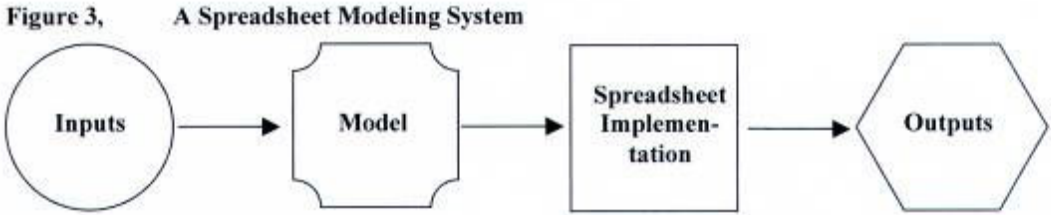

A spreadsheet modelling system is a special case of an information system, where the purpose of the system is to perform the computations in a quantitative model, and spreadsheet software is used to compute the model.

In some cases, model outputs are computed once. In other cases, the user analyses the model by selectively changing inputs and using the model to compute corresponding outputs.

## 3. ACCURACY

We are interested in the *accuracy* of system outputs. **Accuracy** is deviation from the "true" model outputs. Spreadsheet modelling system accuracy is a problematic concept, because we are dealing with models of conceptual processes, as opposed to measurable physical processes.

### 3.1. Evaluating System Accuracy

For modelling systems that predict outputs of measurable physical processes, we can evaluate accuracy by comparing the true measured value to the system outputs. However, in many spreadsheet modelling systems, in can be impossible to objectively determine the "true" outputs: they are *unknowable.* Often, the accuracy of a spreadsheet modelling system is evaluated subjectively. This is in marked contrast to much spreadsheet error research.

There are special cases where we can measure the accuracy of a spreadsheet modelling system. If there is complete agreement on a precise model, and agreement on which input values are to be used, then we can objectively evaluate the accuracy of the implementation, provided we have a guaranteed means of performing accurate computations. This is true for small models. This may also be true for models where a source of authority possesses a model that is believed accurate. For example, taxation authorities have well-tested modelling systems that are capable of generating "true" outputs.

When "true" outputs are unavailable, a proxy is to independently construct two systems and compare the outputs. This seems to be happening for some spreadsheet modelling systems. An example, is that Operis Group PLC's spreadsheet verification service. My understanding is that Operis constructs their own version of a client's spreadsheet model using a proprietary development process for that class of spreadsheets which minimizes (perhaps prevents entirely) spreadsheet implementation errors. They use the client's data to generate outputs equivalent to the client's model's outputs. Thus, the Operis model is considered to generate "true" outputs. By comparing the outputs of the Operis model to the outputs of the client's model, errors in the outputs of the clients model can be identified.

This multi-model approach has inherent limitations. Any errors that exist in both models will not be detected. In addition, error detection is limited to those errors that appear *for that set of data,* so this approach does not guarantee that accuracy is robust in the data.

### 3.2. Standards of Accuracy

It is important to ask the question: What standard of accuracy is appropriate for the outputs of a spreadsheet modelling system? The existing research on spreadsheet errors uses, in essence, a standard of perfect accuracy. This standard of perfection is appropriate when evaluating the quality of a spreadsheet implementation in isolation. Perfection may be an appropriate standard in certain business situations, such as the computation of VAT payments.

However, the standard of perfection may not be appropriate for a spreadsheet modelling system for three reasons: purpose of the modelling system, problems with inputs, and resource constraints.

**Purpose of Modelling System**

The standard of accuracy depends on the purpose of the modelling system. There are many situations where inaccurate results are of great practical benefit. Many exploratory models in business intend only "rough" analysis. There are many business situations where managers have only "gut feel" to guide their actions. Although accurate model outputs would be a valuable contribution, even inaccurate model outputs would be an improvement over the alternative of no analysis at all. For example, [Sonntag and Grossman 1999] discuss a business situation where insights from an analysis of the spreadsheet modelling system had dramatic impact on the business, but where even severe inaccuracies in the spreadsheet implementation would likely have had little effect on the outcome.

Models that are analyzed to obtain broad insight into alternative courses of action can tolerate large inaccuracies. Models that are concerned with achievement of a target can tolerate sizable inaccuracies. For example, when estimating the internal rate of return of an investment the desired result might be a determination that the IRR is well below, about the same as, or well above some hurdle rate.

**Problems with Inputs**

Spreadsheet models are sometimes created for situations where the inputs are known to be of wretched quality. Achieving accurate outputs is essentially impossible. However, the outputs themselves may provide useful insight into important business issues. In addition, these outputs can be analyzed to estimate the value of higher quality inputs.

**Resource Constraints**

The standard of accuracy is limited by the resources available. Often a quick or low-cost analysis at less than the best possible accuracy is valuable, but a perfect or "very accurate" analysis is either impossible, prohibitively expensive, or would be too late to be useful. In such situations, a less-than-perfect standard of accuracy is required.

## 4. SOURCES OF INACCURACY

Inaccuracy in spreadsheet modelling systems can in occur in any of the first three components of Figure 3: inputs, the model itself, and the spreadsheet implementation. See Figure 4.

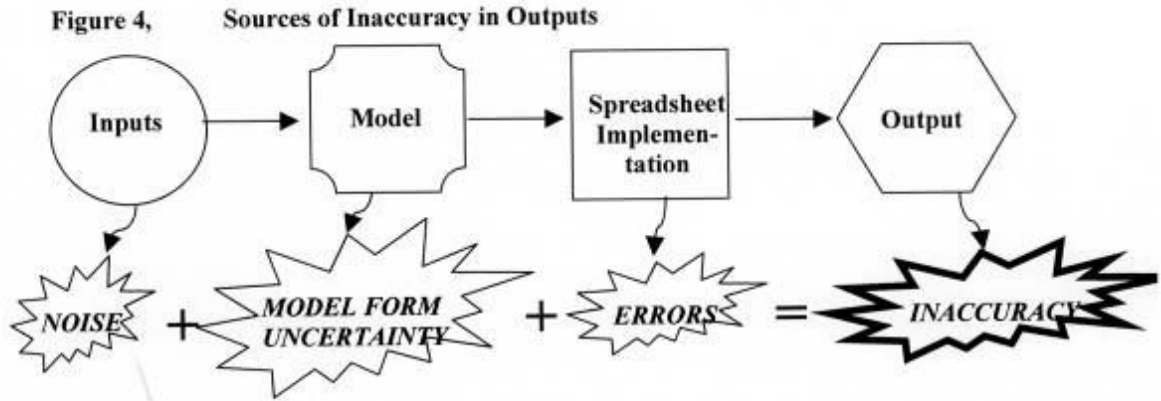

Figure 4, Sources of Inaccuracy in Outputs

### 4.1. Inaccuracy in Inputs: Noise

It is well known that databases are full of incorrect numbers, typographical mistakes, insufficient granularity, and plain nonsense. Thus, in business modelling, inputs generally contain inaccuracies.

In addition, many business models include inputs with inherent random variability. For example, forecasts (of sales, interest rates, and R&D outcomes) are all probabilistic in nature. In principle, such inputs should be modelled as probability distributions. In practice, a modelling choice is often made to use point estimates to approximate probability distributions and avoid concomitant probabilistic analysis. This introduces inaccuracies into the model.

I use the term **noise** to include all inaccuracies in model inputs, whether they are inherent in the data, or occur as a result of a modelling decision.

In most business modelling situations, input noise can be reduced by investing resources over time. A user chooses the amount of effort to expend detecting, evaluating, and ameliorating inaccuracies. Possible actions include; acquire data directly, inquire as to more detailed data, seek to modify data acquisition to provide additional granularity, consult with experts, or replace point estimates with probability distributions.

Noise in model inputs can be handled during model analysis by applying sensitivity analysis techniques from Decision Analysis [Clemen 1996], or by using Monte Carlo simulation [Ragsdale 2001]. The choice of sensitivity ranges and sometimes the choice of probability distributions are subjective.

## 4.2. Inaccuracy in the Model: Model Form Uncertainty

Modelling is inherently subjective. Given the same business situation, different businesspeople will create different models. For example, in modelling courses offered at Stanford University and the Tuck School at Dartmouth College, faculty routinely see a wide range of models promulgated for even simple modelling problems.

Borrowing a term from risk analysis [Vose 2000, Yoe 1996, Agarwal et al 2002, Oberkampf et al 2002], I use **model form uncertainty** to mean differences in the choice of model, degree of simplification in the model, imperfect models, and subjective judgment in modelling including approximations.

For models of the physical world, inaccuracy from model form uncertainty (actually, for model form uncertainty plus implementation error) can be evaluated by comparing model outputs to physical measurements. This is usually not possible for business models, which describe a conceptual world. Thus, it is difficult to objectively evaluate model form uncertainty. It is possible to perform structural sensitivity analysis to compare alternative model forms. For example, see [Caulkins 2001].

Model form uncertainty can be an important source of inaccuracy in outputs. For business situations that are speculative or not well understood, model form uncertainty is likely to be an important source of inaccuracy.

## 4.3. Inaccuracy in the Spreadsheet Implementation: Errors

A spreadsheet computer program implements a model. Spreadsheet research has taught us much about the frequency of errors in spreadsheet implementations. I use the term **error** to refer to outputs from a spreadsheet implementation that is different from the correct model outputs. Research on errors is summarised by [Panko 2000a]

It the "true" model outputs are know, the accuracy of a spreadsheet implementation can be determined by comparing the outputs to the "true" outputs. If the "true" outputs are not known, then we can apply (imperfect) audits to the spreadsheet to detect errors.

It is likely that the process used to develop the spreadsheet affects the error rate. [Grossman 2002] argues that application of sound spreadsheet engineering practices will reduce the error rate, but points out that commingling the act modelling and the act of spreadsheet programming can be a source of errors.

## 5. MANAGING ACCURACY IN SPREADSHEET MODELING SYSTEMS

Figure 5 describes the process of creating and using a spreadsheet modelling system. The user obtains inputs, creates a model, and implements it in a spreadsheet. Because model development is usually iterative [Willemain 1994a 1994b], the user goes through several cycles of examining system components (inputs, model and spreadsheet implementation). He uses judgmental evaluation of the outputs and each of the components to guide allocation of resources at each iteration. The level of inaccuracy from input noise, model form uncertainty, and spreadsheet implementation error are affected by the resources devoted respectively to model inputs, the model itself, and engineering the spreadsheet implementation.

In addition, in cases the user must devote effort employing the spreadsheet modelling system for analysis, interpretation, sensitivity analysis, and creation of reports, charts and tables that communicate model insights. These activities are collectively represented by the box "Business Analysis Using the Spreadsheet Modelling System" in Figure 5. Even with a perfectly accurate spreadsheet modelling system, failure to perform adequate analysis may lead to incorrect or inadequate conclusions from the modelling system.

In practice, virtually all spreadsheet modelling systems are created subject to resource constraints. There are limited resources including money, chronological time, user/manager attention, and social/political considerations. The user must make tradeoffs regarding the resources devoted to each component of the system, and to analyzing the system once it is completed.

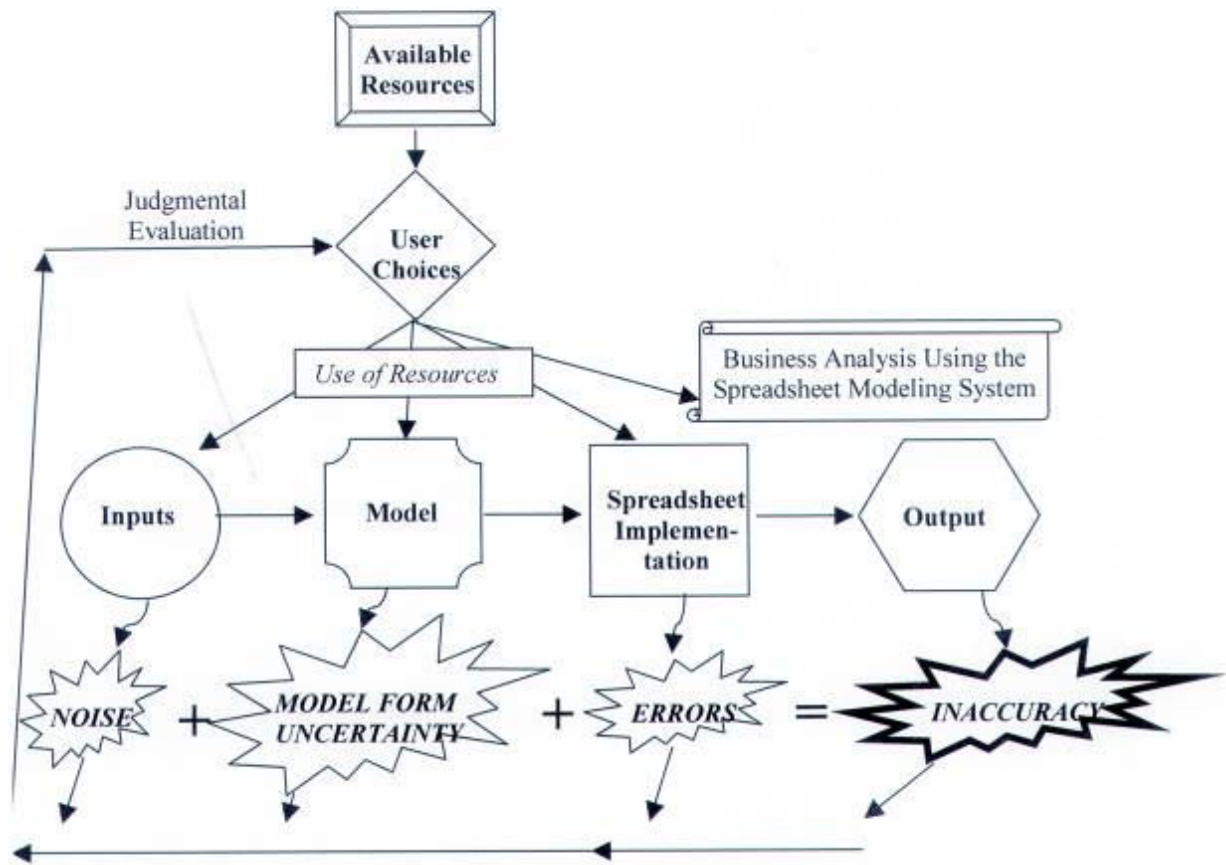

Figure 5, User Choices for a Spreadsheet Modeling System

The user's choices on resource allocation should balance the inaccuracy from each of the three sources, while allowing adequate resources to analyze the system to obtain business insight. Expending excessive resources on one component can lead to problems in others. For an extreme example, common among beginner modellers, some model development efforts focus excessive resources on data development and other activities to acquire and minimize noise in model inputs. This leaves insufficient resources to develop a model that can meaningfully process those inputs, and can waste resources creating high-quality inputs that later analysis determines are not actually needed. The result can be a marvellous database, but little in the way of useful outputs.

On the other extreme, obsessive effort to create a perfect, error-free spreadsheet can consume resources that might have been better spent enhancing the quality of the inputs or improving the quality of the model itself. The result is a spreadsheet that perfectly processes low-quality inputs through an inadequate model, leading to exact computation of low-quality, inaccurate outputs.

## 6. BIASES IN SPREADSHEET MODELLING SYSTEM DEVELOPMENT

During development of a spreadsheet modelling system, the user needs to evaluate noise, model form uncertainty, and errors, as well as the accuracy of the resulting outputs. This evaluation is judgmental, and subject to biases.

We know little about how this evaluation is performed. Based on observation and discussion with students in classroom exercises, conversations with students who formerly developed or supervised important spreadsheets in investment banking and consulting, and conversations with active spreadsheet users, I believe that many spreadsheet users use the *perceived accuracy* of model outputs as their primary evaluation technique.

**Perceived accuracy** of model outputs is how close the outputs are to the outputs the user *expects* to see. If perceived accuracy is high, the user is more satisfied and confident than if perceived accuracy is low.

### 6.1. Expectation Bias

When perceived accuracy is low, the user is likely to devote resources to investigating the system to understand the source of the perceived inaccuracy. ("It's not what I expected, there must be a mistake.") When perceived accuracy is high, the user is likely to devote resources to other activities, particularly analysis using the system, but also model extensions, documentation, or other responsibilities.

The bias occurs because the trigger for investigating the modelling system is the *unexpected* result. This suggests that modelling systems that deliver expected results systematically receive less effort to confirm their accuracy than models that do not. I call this **expectation bias.**

This type of bias is known elsewhere. [Armstrong 1997] discusses the difficulty in publishing research findings that conflict with reviewer's bias in favour of expectations'. [Bazerman el al 2002] describe biases in accounting audits. In laboratory experiments, subjects demonstrate that their judgment in accounting issues is biased toward the interests of the audit client. This is analogous to a user whose judgment is biased toward an expected or desired set of results.

### 6.2. Diagnosis Bias

Low perceived accuracy is caused by two things: either a *mistake,* or a *business insight.* It can be difficult to distinguish the two. A mistake is caused by a problem with the inputs, the model, or the spreadsheet implementation. This problem needs to be identified and fixed.

In contrast, a business insight occurs when unanticipated results are observed and the standard of accuracy is satisfied. The unanticipated model outputs can teach the user something he did not know about the business problem being modelled. In this case, we say that the modelling system is providing business insight. These business insights are an important benefit (sometimes the primary benefit see [Geoffrion 1976]) of building modelling systems.

Ideally, a user will investigate a perceived inaccuracy in an objective, scientific manner, seeking to determine whether the source of the perceived inaccuracy is a mistake or a business insight. In practice, this may not be the case.

Little is known about how spreadsheet model users investigate perceived inaccuracies. In my experience, the user's a *priori* diagnosis of the cause of the perceived inaccuracy affects his actions. The user can form a diagnosis that the outputs are incorrect. Or he can form a diagnosis that the outputs are correct. If the user's actions in response to a perceived inaccuracy are different depending on this diagnosis, he suffers from **diagnosis bias.**

The symptom of diagnosis bias is that the user goes looking for what he thinks he will find. If thinks there is a business insight, he seeks to understand and articulate it. If he thinks there is a mistake, he seeks to find and fix the mistake wherever it may be.

If the user believes a *priori* that there is a business insight, there is a risk he will seek to articulate that insight without adequate evaluation of the accuracy of the modelling system. This risk is particularly high for end-user modellers working under time pressure. Thus, diagnosis bias can lead to false business insights that are actually mistakes in the modelling system.

If the user believes a *priori*, there is a mistake, and then he will begin debugging the system. There is a risk of debugging until the mistake is fixed, rather than debugging until the system is verified. This can take the form of debugging the inputs, model, or spreadsheet implementation until a change is made that shrinks the perceived inaccuracy to an acceptable level. Then victory is declared and the debugging process stops. This process can work effectively if there is but a single mistake. However, the user can introduce a new mistake, or he can fix one mistake but miss other mistakes.

Sometimes, the debugging process will convince the user that the model is correct, and there is in fact an insight to be articulated. He can than work to articulate the insight.

### 6.3. Minimizing the Effects of Biases

The software engineering literature tells us that reliable error detection requires use of structured verification techniques (for example, [McConnell 1993]). [Panko 1999 2000c] makes a similar argument for spreadsheets. In the absence of such verification techniques, it is difficult to evaluate the accuracy of model outputs. The solution to expectation bias is to follow verification procedures that are independent of the perceived accuracy of model outputs. This requires some level of formality in development procedures.

To minimize the effects of diagnosis bias, the user needs to confirm adequate system accuracy before seeking to articulate business insight. This requires some means for verifying that the accuracy of the spreadsheet modelling system is satisfactory. This is no easy task, particularly when the standard of accuracy for the system is low.

### 7. CONCLUSIONS

Inaccuracy in spreadsheet modelling systems can occur due to spreadsheet implementation errors. However, it can also occur due to problems with model inputs, or the model itself. When analysis of the system is required, inadequate time for analysis can lead to inadequate business insights.

## 7.1. Interpreting Existing Research Results

Existing spreadsheet error research tends to focus exclusively on spreadsheet implementation errors. Laboratory spreadsheet error research isolates spreadsheet implementation by controlling for data noise and model form uncertainty. This research provides "perfect" noiseless data. It controls for model form uncertainty by providing a precise problem statement. In fact, the best laboratory research [Panko and Sprague 1999] is careful to eliminate model form uncertainty by providing a "domain free" problem statement.

Field research on errors is summarized in [Panko 2000b]. In general, the field research performs (imperfect) audits for spreadsheet implementation accuracy. This presupposes that the purpose of the spreadsheet modelling system is to generate error-free spreadsheet outputs. However, the purpose of the system is to generate model outputs with accuracy appropriate for the business situation and the resource constraints. Such audits do not consider inaccuracies due to input noise or model form uncertainty. Therefore, this research is most relevant when the spreadsheet implementation is the dominant source of inaccuracy. When this is not the case, this approach misses important sources of inaccuracy.

## 7.2. Understanding Lack of Industry Response to Error Research

Resource constraints on development mean that inaccuracy in system outputs are often unavoidable, and industry has presumably learned to cope with them. Indeed, it is only in living memory that managers have had access to any quantitative analysis at all. Since an analysis with limited accuracy is generally better than the alternative of no analysis at all, imperfect accuracy may not be a serious concern in many circumstances. This suggests that inaccuracies due to spreadsheet errors may be tolerable in practice.

The argument that spreadsheet errors is a serious problem is most valid in those situations where spreadsheet implementation errors are the main source of inaccuracy, and where the inaccuracy is of sufficient magnitude to be painful. Such situations certainly exist, but they may not be common. Research on the prevalence of such special cases is needed.

Efficiencies in managing the accuracy of any of the sources of system inaccuracy can have valuable knock-on effects. For example, if new practices can reduce the resources required to construct a spreadsheet for a given level of accuracy, the resources saved could be used to enhance the quality of the model inputs or the model itself, or to perform additional analysis. Thus, <u>spreadsheet research should consider the productivity of spreadsheet construction</u> as well as the prevention of errors.

**ACKNOWLEDGMENTS**

Thanks to Stephen Powell and Ralph Scheubrein for their comments on a draft.